\documentclass[aps,pre,reprint,superscriptaddress]{revtex4-1}

\usepackage{graphicx}
\usepackage{dcolumn}
\usepackage{bm}

\begin{document}


\title{Integrated simulation approach for laser-driven fast ignition}

\author{W.-M. Wang}
\email{Please also see this paper in Phys. Rev. E 91, 013101 (2015)}
\affiliation{Forschungzentrum Juelich GmbH, Institute for Advanced
Simulation, Juelich Supercomputing Centre, D-52425 Juelich, Germany}
\affiliation{Beijing National Laboratory for Condensed Matter
Physics, Institute of Physics, CAS, Beijing 100190, China}
\author{P. Gibbon}
\affiliation{Forschungzentrum Juelich GmbH, Institute for Advanced
Simulation, Juelich Supercomputing Centre, D-52425 Juelich, Germany}
\author{Z.-M. Sheng}
\affiliation{Department of Physics, SUPA, Strathclyde University,
Rottenrow 107, G4 0NG Glasgow, United Kingdom} \affiliation{Key
Laboratory for Laser Plasmas (MoE) and Department of Physics and
Astronomy, Shanghai Jiao Tong University, Shanghai 200240, China}
\author{Y.-T. Li}
\affiliation{Beijing National Laboratory for Condensed Matter
Physics, Institute of Physics, CAS, Beijing 100190, China}

\date{\today}

\begin{abstract}
An integrated simulation approach fully based upon particle-in-cell
(PIC) model is proposed, which involves both fast particle
generation via laser solid-density plasma interaction and transport
and energy deposition of the particles in extremely high density
plasma. It is realized by introducing two independent systems in a
simulation, where the fast particle generation is simulated by a
full PIC system and the transport and energy deposition computed by
a second PIC system with a reduced field solver. Data of the fast
particles generated in the full PIC system are copied to the reduced
PIC system in real time as the fast particle source. Unlike a
two-region approach, which takes a single PIC system and two field
solvers in two plasma density regions, respectively, the present one
need not match the field-solvers since the reduced field solver and
the full solver adopted respectively in the two systems are
independent. A simulation case is presented, which demonstrates that
this approach can be applied to integrated simulation of fast
ignition with real target densities, e.g., 300 $\rm g/cm^3$.
\end{abstract}


\maketitle

\section{Introduction}
As an alternative route to realize laser fusion energy, the fast
ignition (FI) scheme has attracted significant attention since it
was proposed by Tabak {\it et al.} two decades ago \cite{tabak}.
This scheme separates the compression and ignition processes and
consequently relaxes the constraints of compression density and
symmetry, which may offer the possibility of higher energy gain. A
high coupling of energy has been demonstrated experimentally by
Kodama {\it et al.} in 2001 \cite{kodama}. In their experiment a
metal cone is embed into the target to guide the propagation of fast
electrons and the ignition laser as well as to reduce the distance
of the electrons to the target core. Since then, many groups
worldwide have performed studies on all aspects of FI physics
including generation of fast electrons via laser plasma interaction,
transport of the electrons in coronal plasma with steep density
gradient, and heating of the target core \cite{sheng,chen,lei,li,
Debayle,Cui,santos,matsumoto,li2,sentoku,kemp,mason,honrubia,Atzeni}.
Most studies have mainly focused upon one or two of these processes.

Integrated investigations of these processes are essential to fully
assess the FI scheme. For this purpose, integrated experimental
studies with large-scale FI targets are ideal and play a decisive
role, however which need high energy laser beams and carefully
designed targets, involving difficulties in both technology and
physical understanding. Prior to such experiments, large-scale,
integrated numerical simulations are a good choice. But it is hard
to find a model to describe all the involved processes. Traditional
particle-in-cell (PIC) models \cite{PIC1,PIC2,Yee_solver} are
suitable to simulate the generation of fast electrons from
interaction of laser and plasma below and around 100 $n_c$, where
$n_c=1.1\times 10^{21} \rm cm^{-3}$ is the critical density
corresponding to 1-$\mu m$-wavelength lasers. It is almost
impossible to apply this model to the problem of fast electron
transport in FI plasma targets with the density growing from
hundreds of $n_c$ to tens thousands of $n_c$ due to huge numerical
noise appearing with unresolved plasma oscillation. A hybrid PIC
model \cite{hybrid1,hybrid2,hybrid3,hybrid4} has been employed to
calculate the electron transport and energy deposition, in which the
background plasma is considered as a fluid. In this case, the
displacement current in Ampere's law is omitted and high frequency
dynamics of the background plasma need not be resolved. Hence, this
approach cannot describe laser plasma interaction.

To include all the three processes within an integrated model,
Sentoku and Kemp proposed to artificially reduce the plasma density
in collisional PIC simulation when it exceeds a upper-limit value,
e.g., 500 $n_c$ \cite{Sentoku_08}. In this approach a macroparticle
has two weights: one is its real weight which is used to calculate
Coulomb collision and another is a reduced weight to calculate the
current for the field solver. Hence, the electromagnetic (EM) fields
are not consistent with the real plasma density, which can be
applied when the resistive effect is much less than the collisions.
To use this approach, one should make sure that the numerical noise
at high density region is controlled to be quite low since the
energy of a macroparticle with a reduced weight gaining from the
noise is amplified by the ratio of the real weight to the reduced
one. With this model Chrisman {\it et al.} have performed a group of
integrated FI simulations with the core density as high as
$20000~n_c$ or 100 $\rm g/cm^3$ \cite{chrisman}.

An improved model, named the two-region PIC, was proposed by Cohen
{\it et al.} \cite{Cohen} in 2010, in which the high plasma density
is not clamped artificially. In this model, the simulation box is
separated into a low plasma density region and a high density
region, where the density at the boundary between the two regions is
taken as $\sim 100~n_c$. In the low density region a full PIC
algorithm with collisions is taken and in the other region the
Maxwell's equations are reduced by use of the Ohm's law to solve the
electric fields while the Ampere's law is used to calculate currents
of background electrons. This reduced field solver is similar to the
one used in the hybrid PIC model
\cite{hybrid1,hybrid2,hybrid3,hybrid4}, whereas the background
plasma comprises macroparticles as in a traditional PIC model. In
this case the EM fields are consistent with the plasma density in
the whole simulation region. However, a potential challenge for this
model arises in that the continuity of EM fields near the boundary
of the two regions can be violated due to the noise of the full
field solver. This noise is usually several orders of magnitude
higher than the one of the reduced field solver, which may mask the
real value given by the reduced field solver after a long period of
simulation. Therefore, matching the EM fields around the boundary
becomes challenging, although it may be partially solved by
increasing the spatial resolution and using a large number of
macroparticles per cell.

Here, we propose an approach to release the constraint of the EM
field matching around the boundary of the two regions. Our approach
involves two independent simulation systems: one using the full PIC
model and another applying the reduced model as adopted before in
the high density region of the two-region PIC approach \cite{Cohen}.
Moreover both the full and reduced Maxwell's equations are solved in
the whole simulation box, and the two solutions are independent in
our approach. The full PIC system is used to simulate the generation
of fast particles via laser plasma interactions. Data of the
generated fast particles are copied to the reduced-field-solver PIC
system in real time but retained in the full PIC system. The
transport of the particles and energy deposition are calculated in
the reduced-field-solver PIC system. In this way there is no longer
any need to match the two field solutions as usually required in a
single PIC system with a boundary to separate two field solvers. The
resolution in both PIC systems in our approach can be taken as an
usual one to satisfy the full field solver for certain maximum
plasma density (e.g. $100~n_c$), since the reduced field solver need
not a high resolution. We call this approach two-system PIC.

The outline of the paper is as follows. In Sec. II, a comparison is
made about the noise between the full PIC simulation and the
two-region PIC simulation, which also provides benchmark for the
reduced field solver. In Sec. III, the basic algorithm for the
two-system approach is presented. An example of integrated FI
simulation with the two-system approach is given in Sec. IV. We
conclude with a summary and discussion in Sec. V.

\section{Comparison between the reduced and full field solvers}
\begin{figure}[htbp]
\includegraphics[width=3.6in,height=3.6in]{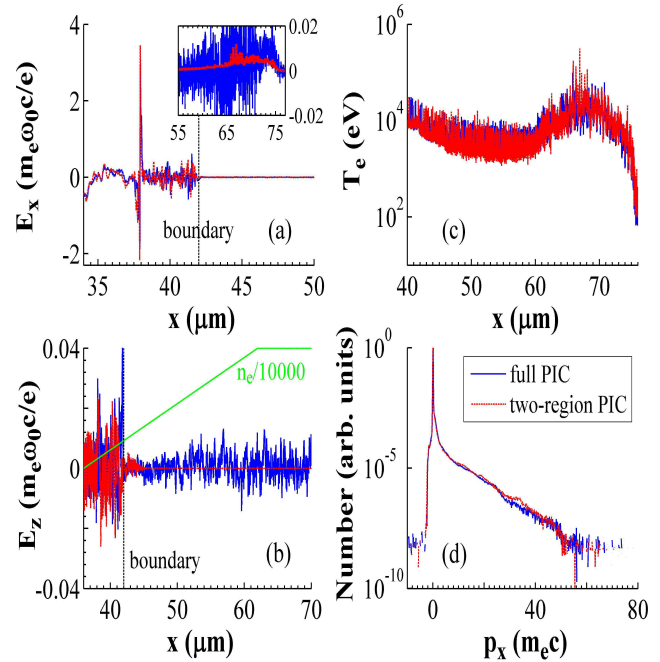}
\caption{\label{fig:epsart} Comparison of 2D simulation results
obtained with the full PIC simulation (blue-solid lines) and the
two-region PIC simulation (red-broken lines). (a)-(c) Spatial
distributions of electric fields and electron temperature at the
axis (y=0). The inset in (a) is a close-up of the field. (d)
Electron number distribution as a function of the longitudinal
momenta. These results are given at 0.15 ps. }
\end{figure}

As mentioned in the introduction, our two-system PIC approach is
based upon the two-region PIC model, where a reduced field solver is
adopted for the high density region. The two-region PIC model has
been benchmarked fully against the full PIC model in both 1D and 2D
geometry in Ref. \cite{Cohen}. Here we present an extended benchmark
in 2D geometry with a higher spatial resolution and a larger number
of macroparticles to display the matching problem of the full and
reduced field solvers on the boundary. In our simulations, the
spatial resolution in both the x and y directions is 5 nm. A cell is
loaded with 400 macroparticles for each species. Figure 1 shows our
benchmark simulation with our PIC code KLAPS (see the introduction
of KLAPS in Appendix A) and illustrates the reduced field solver can
be applied in high density plasma region, as shown in Ref.
\cite{Cohen}. The simulation setup is as follows. The plasma density
grows linearly from 0.2 $n_c$ to 400 $n_c$ between $x=36~\mu m$ and
$x=62~\mu m$ and remains at a plateau of 400 $n_c$ with a size of
$14~\mu m$. A laser pulse at wavelength 1 $\mu m$ propagates along
the +x direction with peak intensity of $5\times 10^{19}~Wcm^{-2}$,
linear polarization along the y direction, and a full width at half
maximum (FWHM) duration of 30 fs. To reduce the computation expense
we take a planar laser profile so that the simulation box size along
the y direction can be set small enough. In the two-region PIC
simulation, the boundary at $x=42~\mu m$ is taken, as illustrated in
Figs. 1(a) and 1(c). The fields are given by the full field solver
(solving the full Maxwell's equations as in a traditional PIC model
\cite{PIC1,Yee_solver}) in the left region with the plasma density
below 100 $n_c$ and those are solved by the reduced Maxwell's
equations \cite{Cohen} in the right region with the plasma density
above 100 $n_c$. In the full simulation there is no such a boundary
and the fields in the whole simulation box are obtained by the full
field solver. In both simulations, a Coulomb collision module and a
4th order current calculation are employed (see Appendix A) and the
digital smoothing of fields around the boundary and high density
region is taken as Ref. \cite{Cohen} (although there is no boundary
in the full PIC simulation).

Here, the resolution and particle number per cell are taken to be
much higher than those in usual simulations. However, the numerical
noises of the fields in the full PIC simulation still appear to be
much higher than those in the right region of the two-region PIC
simulation, as clearly seen in Figs. 1(a) and 1(c). The noise of the
former is 2-3 orders of magnitude higher than the latter, which will
grow further if a lower resolution is taken. In particular, the
noise of $E_z$ in the left region in the two-region PIC simulation
is also rather higher than the one in the right region (note $E_z$
should vanish in theory since the laser polarization is along the y
direction). One notices that the noise of the full field solution in
the left region has spread to $x=46~\mu m$ (the boundary at
$x=42~\mu m$) at this time. The value given by the reduced field
solver is masked, which can result in the discontinuity of EM fields
around the boundary of the two regions. This may limit the
application of the two-region PIC approach if a long time's
simulation is required (e.g., for FI simulation) and at the
meanwhile a moderate resolution is employed due to limited
computational resources.

\section{Algorithm for the two-system PIC approach}
Our two-system PIC approach can completely avoid the problem of
field discontinuities close to the boundary, even with a
conventional resolution. This approach needs only a few
modifications to the two-region PIC approach. We solve both the full
and reduced Maxwell's equations in the \textit{whole} simulation box
independently in two systems. The two systems have their own
respective macroparticles, which are independent of each other. We
denote the full PIC system as system I and the reduced-field-solver
PIC system as system II. In system I a traditional PIC algorithm
with Coulomb collisions \cite{Sentoku_08} is taken to simulate the
generation of fast particles via laser plasma interactions. Because
the ignition laser pulse cannot propagate into a region with high
density plasma in a FI case, we lower artificially the density in
this region to a given value when it exceeds this value (e.g.,
hundreds of $n_c$). Before the generated fast particles enter this
region, their data (positions, momenta, masses, charges, weights,
etc.) in system I are duplicated to system II in real time (and
meanwhile these particles are retained in system I). In system II,
the plasma density is taken as the real profile in the whole
simulation box and the fields are solved with the reduced Maxwell's
equations, similar to the ones used in the high density region in
the two-region PIC approach \cite{Cohen}. Note that the incident
laser pulse is found only in system I. Obviously, there is no need
to match the full and reduced field solutions in the two-system
approach.

The duplication of the fast particle data is performed in the
following way. In system I, one calculates the generation of fast
particles and the transport to a region which is far enough away
from interacting points of the laser. In the region, the data of the
fast particles are copied to system II, where we call the region as
the injection point of the fast particles. One can artificially
reduce the plasma density from its real value behind the injection
point to reduce numerical noise in system I, as mentioned above. In
system II the plasma density is not changed artificially. The setup
of the density and the injection point can be seen in Fig. 2 below
as an example.

We present the algorithm equations of the field solver in system II
(note that the full Maxwell's equations are taken in system I). The
background currents of electrons and ions are calculated by the
Ampere's law omitting the displacement current,
\begin{eqnarray}
\textbf{J}_b=\frac{c}{4\pi} \nabla \times \textbf{B}-\textbf{J}_f,
\end{eqnarray}
where $\textbf{J}_f$ is the total current of fast electrons and
ions, which is computed from the PIC particles directly. The
background current calculated in this way avoids numerical noise in
extremely high density, e.g., $300g/cm^3$. Faraday's law is used to
advance the magnetic fields:
\begin{eqnarray}
\frac{\partial \textbf{B}}{\partial t}=-c \nabla \times \textbf{E}.
\end{eqnarray}
The electric fields are solved by the Ohm's law:
\begin{eqnarray}
\textbf{E}=\bm{\eta} \cdot \textbf{J}_b,
\end{eqnarray}
where $\bm{\eta}$ is the classical resistivity
\cite{Cohen,Braginskii}. The field solver given by Eqs. (1)-(3) is
the same with the one used in hybrid PIC model
\cite{hybrid1,hybrid2,hybrid3,hybrid4}, the validation of which has
been established. The largest difference between the hybrid PIC
model and our system II is that the former considers the background
plasma as a fluid denoting by $\textbf{E}=\bm{\eta} \cdot
\textbf{J}_b$, whereas the latter takes the background as
macroparticles as in a traditional PIC model. In principle, the
calculation of the collisions between fast particles and the
background plasma is more accurate in system II than that in hybrid
PIC simulation since it uses a series of Monte Carlo tests between
randomly-chosen particle pairs computed with the relativistic binary
collision formula \cite{Sentoku_08}.

The main difference of the field solver by Eqs. (1)-(3) from the
ones in the two-region PIC approach is that the latter retains the
displacement current term in Eq. (1) to satisfy the numerical
stability constraints \cite{Cohen}. Our simulation shows that such
instability arises when the fluid quantities such as the electron
density and electron average velocity are computed directly from the
particles. This can be canceled in the following way even without
the displacement current term. According to Eq. (1), one can easily
obtain $\nabla \cdot\textbf{J}=0$, where
$\textbf{J}=\textbf{J}_b+\textbf{J}_f$ is the total current.
Applying the continuity equation $\nabla \cdot\textbf{J}+\partial
\rho /\partial t=0$ of the fluid, one gets the charge conservation
$\rho \equiv 0$ all the time if it meets initially. Here $\rho$ is
the total charge density and it can be written by $\rho=eZ n_i-e
n_e-e n_f$, where $n_e$ and $n_i$ are densities of background
electrons and ions and $n_f$ is the fast electron density. Hence,
the background electron density can be given by $n_e=Z n_i-n_f$,
which is consistent with the background current calculated by Eq.
(1). In the case with high density plasma, $n_e \simeq Z n_i$ is a
good approximation. In addition, the charge conservation is met
automatically with the field solver with Eqs. (1)-(3) and one need
not carry out the charge correction.

In the KLAPS code Eqs. (1)-(3) are computed as follows. Fast
particle current $\textbf{J}_f$ as well as $\textbf{J}_b$ and the
temperature of background electrons $T_e$ are defined at
half-integer time points. Fields, densities of background electrons
$n_e$, ions $n_i$, and fast electron density $n_{f,e}$, are defined
at integer time points. Quantities $\textbf{J}_f$, $T_e$, $n_i$, and
$n_{f,e}$ are computed from the particles directly and the others
are derived. Here $T_e$, $n_e$, and $n_i$ are used to calculate the
classical resistivity. The half-integer and integer points in space
are taken according to the traditional Yee scheme \cite{Yee_solver}.
We first calculate the magnetic fields at time $(n+1/2)\triangle t$
using the quantities at $n \triangle t$:
\begin{eqnarray}
\frac{\textbf{B}^{n+1/2}-\textbf{B}^{n}}{\triangle t/2}=-c \nabla
\times \textbf{E}^{n}.
\end{eqnarray}
Then compute the background current at $(n+1/2)\triangle t$:
\begin{eqnarray}
\textbf{J}_b^{n+1/2}=\frac{c \nabla \times \textbf{B}^{n+1/2}}{4\pi}
-\textbf{J}_f^{n+1/2}.
\end{eqnarray}
Insert $\textbf{J}_b^{n+1/2}$ into the Ohm's law and obtain:
\begin{eqnarray}
\textbf{E}^{n+1/2}=\bm{\eta} \cdot \textbf{J}_b^{n+1/2}.
\end{eqnarray}
By use of
\begin{eqnarray}
\textbf{E}^{n+1}=2\textbf{E}^{n+1/2}-\textbf{E}^{n},
\end{eqnarray}
one gets $\textbf{E}^{n+1}$. Finally, calculate the magnetic fields
at $(n+1)\triangle t$ using those at $(n+1/2) \triangle t$:
\begin{eqnarray}
\frac{\textbf{B}^{n+1}-\textbf{B}^{n+1/2}}{\triangle t/2}=-c \nabla
\times \textbf{E}^{n+1}.
\end{eqnarray}

In system II digital smoothing is taken for all fields, currents,
temperatures, and densities. We examine two spatial smoothing
algorithms, one is
\begin{eqnarray}
F_i=\frac{F_{i-1}+2F_i+F_{i+1}}{4},
\end{eqnarray}
and the other is
\begin{eqnarray}
F_i=\frac{F_{i-2}+2F_{i-1}+3F_i+2F_{i+1}+F_{i+2}}{9},
\end{eqnarray}
the subscript $i$ indicates the spatial grid index in either $x$ or
$y$ direction, e.g., for 2D case. This digital smoothing is
performed in both $x$ and $y$ directions. The The two algorithms
give nearly the same results. Considering that 4th-order ``zigzag''
algorithm is employed in both systems (see the introduction of KLAPS
in Appendix A), we take the latter to match 4th-order interpolation
used in both particle pusher and current calculation in KLAPS.

\section{An example of two-system PIC simulation}
\begin{figure}[htbp]
\includegraphics[width=5.0in,height=3.75in]{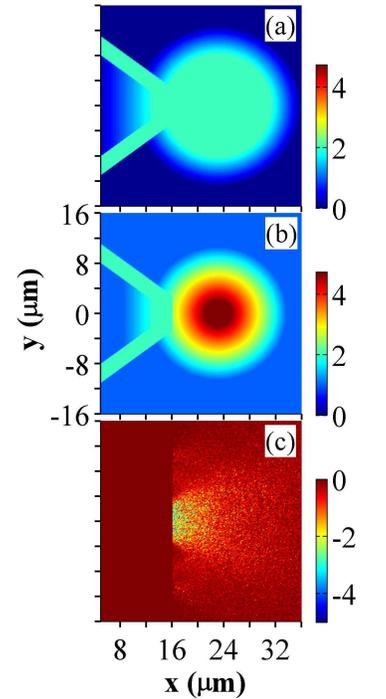}
\caption{\label{fig:epsart}Initial plasma electron densities
$\lg(n_e/n_c)$ taken in system I (a) and system II (b). (c) The fast
particle current $J_{f,x}/en_cc$ in system II at 0.2 ps. The
injection point is taken at $x=16\mu m$.}
\end{figure}

In Figs. 2(a) and 2(b), we demonstrate the density setup in the two
systems through an example of FI. A compressed target is taken with
the uniform density of 300 $\rm g/cm^3$ (or 54000 $n_c$) at the core
area within a circle of the radius 2 $\rm\mu m$ and the surrounding
density decreasing exponentially with a scalelength 1 $\rm\mu m$
along the radial direction away from the core center. A cone is
embed into the target with the cone angle of 45 degree, the wall
depth of 3 $\rm\mu m$, the density of 100 $n_c$, tip size of 4
$\rm\mu m$, and the inner length of 8 $\rm\mu m$. Inside the cone a
preplasma has an exponential profile with a scalelength 2 $\rm\mu m$
along the x direction and is distributed uniformly in the y
direction. System II takes this plasma density profile with a
pedestal of 10 $n_c$, as shown in Fig. 2(a). In system I the density
profile is changed in such a way as when the plasma density is above
the cone wall density 100 $n_c$, it is lowered to be 100 $n_c$, as
displayed in Fig. 2(b). The injection point of the fast particles is
chosen at the cone tip end with $x=16\rm\mu m$. To be related with
FI studies, we define such particles to be fast that the energy is
higher than 0.1 MeV and the forward momentum $p_x>0.45 m_ec$ (50
keV) for electrons or $p_x>46 m_ec$ (0.1 MeV) for tritium ions. The
data of such fast particles are copied from system I to II in real
time.

The two systems have the same simulation box size of $48\rm\mu m
\times 48\rm\mu m$ in $x\times y$ directions. The resolutions in
both the x and y directions are (1/32)$\rm\mu m= 1.96 c/\omega_p$
($\omega_p^2=100n_c\times 4\pi e^2/m_e$) and the temporal resolution
is 0.067fs $=1.25 /\omega_p$. Initially, 25 electrons and tritium
ions cover a cell both in system I and II. The initial temperatures
of electrons and ions are uniformly in space, which is 1 keV. A
laser pulse with wavelength of 1 $\rm\mu m$ propagates along the +x
direction. It is linearly polarized along the y direction with the
electric fields $E=a_0 \exp(-y^2/r_0^2) f(\xi) \sin(2\pi \xi)$,
where $a_0=12.1$ corresponding to $2\times10^{20}\rm W/cm^2$,
$\xi=t-x/c$, $r_0=4~\rm\mu m$, the temporal profile $f(\xi)$ is
taken as a trapezoid, i.e., a plateau of 1 ps between 3.33 fs rising
and decreasing regions. The simulation time is 1.2 ps. Here, these
resolutions and particle number per cell have been examined in
detail as found in Appendix B.

Figure 2(c) shows the longitudinal fast current in system II at 0.2
ps. This fast current is contributed not only by the fast particles
injected from system I but also by the particles originating in
system II which gain high enough energy from the fast particle
influx. We calculate the fast current from the electrons with energy
$E>5T_e(x,y)$ and $E>50$ keV as well as the tritium ions with
$E>5T_i(x,y)$ and $E>50$ keV, where $T_e(x,y)$ and $T_i(x,y)$ are
the local temperatures of electrons and ions.

\begin{figure}[htbp]
\includegraphics[width=3.2in,height=2.8in]{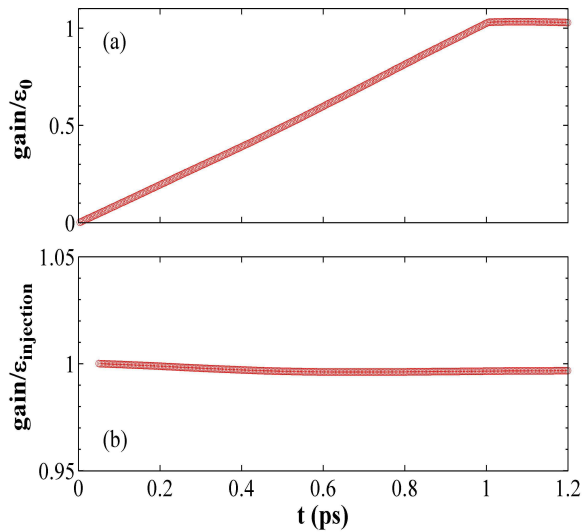}
\caption{\label{fig:epsart}(a) Total energy change with time in
system I, which is normalized by the incident laser energy
$\varepsilon_0$. (b) Total energy change of system II normalized by
the total energy of the fast particles injected
$\varepsilon_{injection}$ at different times. }
\end{figure}

We first check energy conservation of the two systems. Absorbing
boundary conditions for both particles and EM fields are taken in
the two systems and energy of the particles and fields fled away
from the simulation box is recorded. Figure 3(a) displays the energy
gain of system I normalized by the incident laser energy. This
figure illustrates that the energy conservation remains good in
system I. The highest error value in this system is 3.3\% at 1.2 ps.
Figure 3(b) shows the energy of system II which includes the
residual energy of the fast particles, the gain of background
particles, and the EM fields computed by Eqs. (4)-(8). The energy is
normalized by the total energy of the fast particles injected at
different times. The curve of the energy evolution begins at 46 fs
when some fast electrons start to be injected. It shows good energy
conservation within the whole simulation period of 1.2 ps, where the
conservation is kept within 0.4\%. Note that the error for energy
conservation is larger for system I, which is due to the higher
noise of the full field solver. This noise can be lowered by
increasing the number of macroparticles in system I. A benchmark of
our two-system PIC model against the full PIC model is also
presented in Appendix D.

\begin{figure}[htbp]
\includegraphics[width=3.6in,height=3.6in]{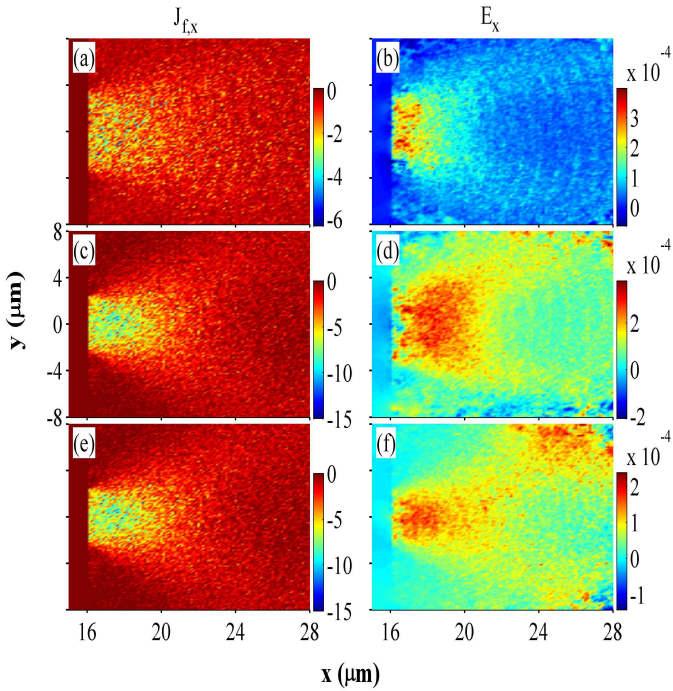}
\caption{\label{fig:epsart}Snapshots of spatial distributions of the
fast current $J_{f,x}$ (the left column) and $E_x$ (the right
column), which are normalized by $en_cc$ and $m_e\omega_0 c/e$,
respectively. The three rows correspond to 0.2 ps, 0.6 ps, and 1 ps.
}
\end{figure}

\begin{figure}[htbp]
\includegraphics[width=3.2in,height=2.8in]{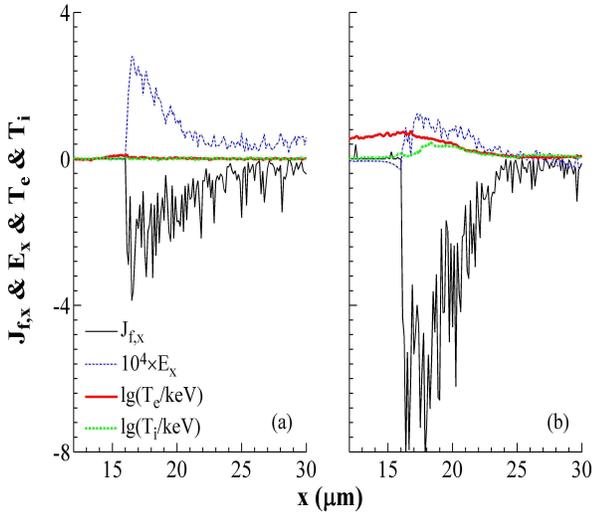}
\caption{\label{fig:epsart}Spatial distributions of $J_{f,x}/en_cc$,
$10^4 \times e E_x /m_e \omega_0 c$, the background temperatures of
electrons $T_e$ and ions $T_i$ on the axis (y=0) at 0.2 ps (a) and
1.2 ps (b).}
\end{figure}

In the following Figs. 4-8 we plot spatial distributions of the EM
fields, currents, and temperatures in system II at different times.
Figure 4 shows the fast currents and longitudinal electric fields at
0.2, 0.6, and 1 ps. Compared to the fast current with the peak at
the injection point in Fig. 4(a) at earlier time, Figs. 4(c) and
4(e) display the peaks shift to the region between the injection
point and the core area beginning at $x=21\rm \mu m$ because a large
number of fast electrons are slowed down by strong collisions and
accumulated at this region. Except with very high energy, fast
electrons are mostly barred from the core and loop around the core
surface. The fast current appears the minimum at the core and the
area just behind it. A similar pattern for the $E_x$ distribution
can be found in Figs. 4(b), 4(d), and 4(f) since $E_{x}=\eta_x
J_{b,x} \simeq - \eta_x J_{f,x}$. A clearer picture can be seen in
Fig. 5. Compared Figs. 5(a) and 5(b), one can observe that the
electric field attenuates with the time although the current is
enhanced. The reason is that $\eta$ strongly depends on the local
electron temperature with $\eta \propto T_e^{-3/2}$. The temperature
grows with time and therefore the electric field attenuates.

\begin{figure}[htbp]
\includegraphics[width=3.6in,height=3.6in]{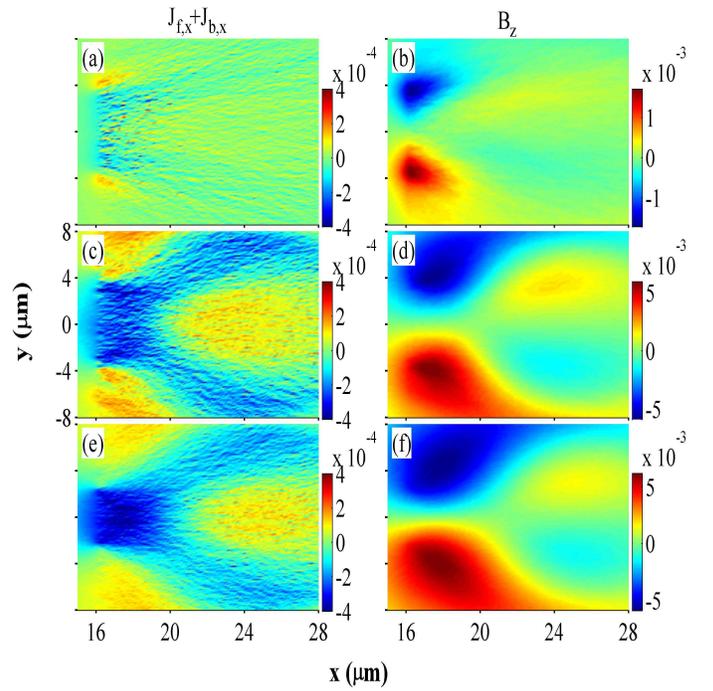}
\caption{\label{fig:epsart}Snapshots of spatial distributions of the
total current along the x direction (the left column) and $B_z$ (the
right column), which are normalized by $en_cc$ and $m_e\omega_0
c/e$, respectively. The three rows correspond to 0.2 ps, 0.6 ps, and
1 ps. }
\end{figure}

The distributions of the total currents (sum of the fast and
background currents) and magnetic fields are illustrated in Fig. 6.
Fine filamentary structures can be seen in the current all the time.
It is also seen that the total currents appear positive sign at the
core area because the return current exceeds the local $J_{f,x}$.
The total current is negative in the surrounding area with lower
densities, where $J_{f,x}$ is not completely neutralized by
$J_{b,x}$. There are also positive total currents in the low density
area just adjacent to and outside the fast current peak area (see
Fig. 4), which are excited by the adjacent fast ones. Accordingly,
the magnetic fields show two groups of peaks with opposite signs:
one around the core and the other around the injection point.

\begin{figure}[htbp]
\includegraphics[width=3.6in,height=3.6in]{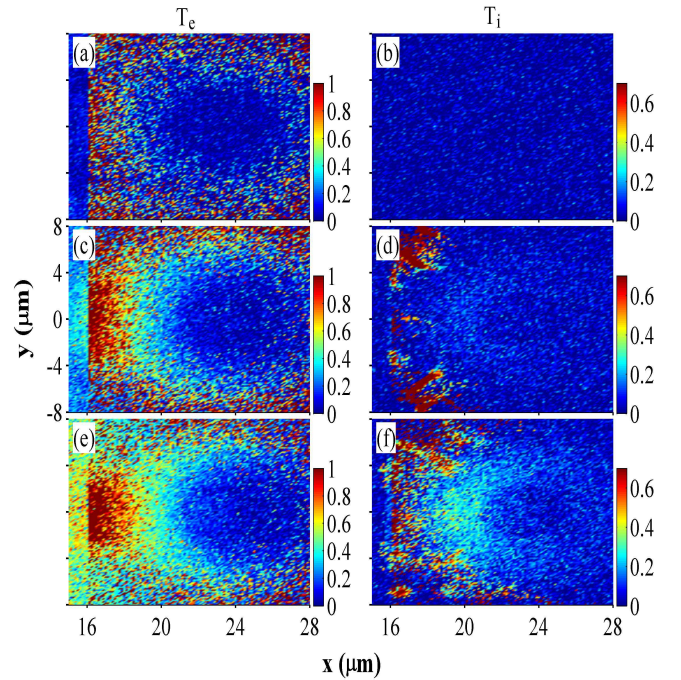}
\caption{\label{fig:epsart}Snapshots of spatial distributions of
$\lg{(T_e/keV)}$ (the left column) and $\lg{(T_i/keV)}$ (the right
column), where $T_e$ and $T_i$ are temperatures of all the electrons
and ions. The three rows correspond to 0.2 ps, 0.6 ps, and 1 ps. }
\end{figure}

\begin{figure}[htbp]
\includegraphics[width=3.6in,height=3.6in]{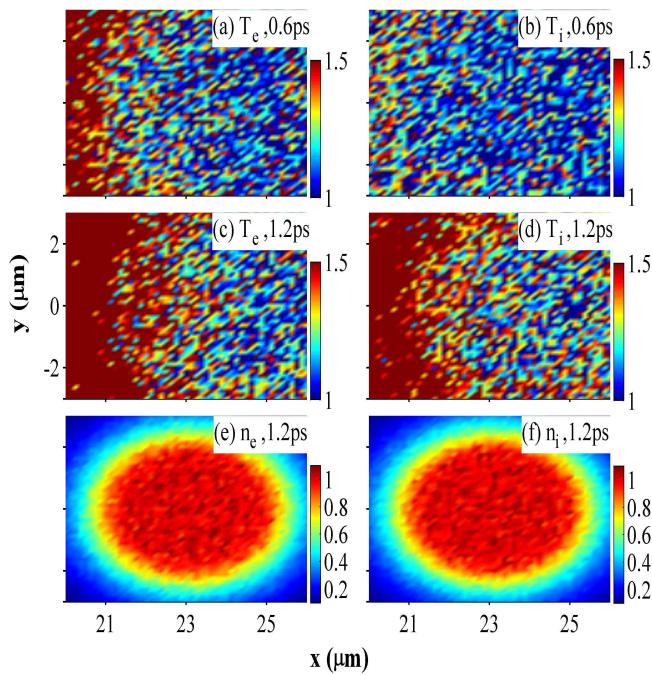}
\caption{\label{fig:epsart}Snapshots of spatial distributions of
$\lg{(T_e/keV)}$ [(a) and (c)] and $\lg{(T_i/keV)}$ [(b) and (d)],
where (a), (b) correspond to 0.6 ps and (c), (d) to 1.2 ps, and
$T_e$ and $T_i$ are temperatures of all the electrons and ions.
Spatial distributions of densities of the electrons and ions around
the core at 1.2 ps are given in (e) and (f), respectively, which are
computed from the particles directly and normalized by 54000 $n_c$.}
\end{figure}

The temperatures of electrons and ions at different times are
plotted in Fig. 7 (the values of the temperatures are shown within a
limited range). One sees that the electron temperature increases
quickly in the area surrounding the core as the fast electrons
arrive (see Fig. 4). The core always remains colder however: between
the core and the injection point a temperature front is observed to
advance slowly towards the core. After some delay, the ion
temperature also increases up to $T_e$ due to collisions. The
temperature peaks in Fig. 7(d) appear at low density regions where
the ions are easy to heat. A clearer picture in Fig. 5 can be seen
that the ion temperature trails with the electron temperature, which
grows with time. One also observes that the electron temperature at
the region in front of the injection point is enhanced at latter
time because some electrons acquire velocities along the -x
direction via collisions and some of these go to the region.

A zoom of the temperatures around the core is given in Figs. 8 (a)-
8(d). One sees a slow process of the electron heating towards the
core. The ion heating with a delay is also observed. One notice that
there is inhomogeneity in the electron and ion temperatures even at
the core with an extremely high density. With such a low
temperature, the rate eliminating the inhomogeneity does not depend
on the collision frequency but the speed of the particles. Figures
8(e) and 8(f) display the densities of the electrons and ions around
the core at 1.2 ps. Deviation of the densities from the initial
values is not large.

\begin{figure}[htbp]
\includegraphics[width=3.2in,height=2.8in]{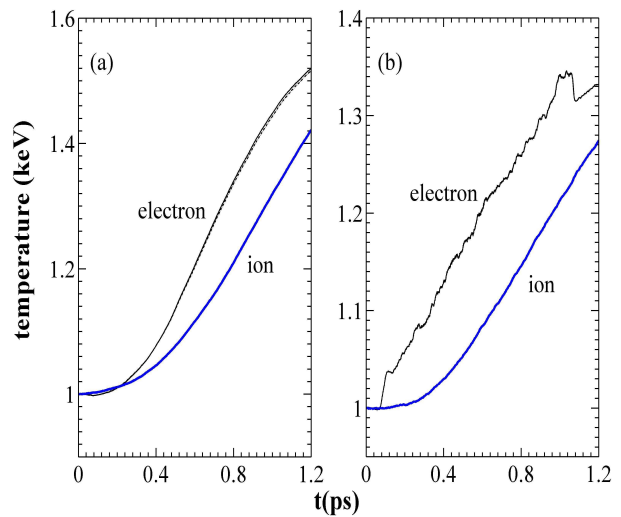}
\caption{\label{fig:epsart}(a) Evolution of the temperatures of the
electrons and ions of system II, where the broken lines denote the
gain from collision and the solid lines are for the total gain. (b)
Evolution of the temperatures of the electrons and ions within the
core area. }
\end{figure}

It is interesting to present the temporal evolution of the
temperatures at the core area in Fig. 9(b). The temperature of the
electrons at the core (including fast electrons just arrived) are
always higher than $T_i$, which leads to the continuous heating to
the ions. The ions are heated to about 1.28 keV at 1.2 ps. Figure
9(a) illustrates the evolution of the temperatures of the electrons
and ions of system II (not including the fast particles injected).
The broken lines mean the gain from collision only while the solid
lines correspond to the total gain. It is shown that the collisional
heating dominates the resistive heating nearly completely. Note that
the latter will be enhanced provided the scalelength of the density
surrounding the core grows, in which the ratio of plasma with lower
density goes up.

\begin{figure}[htbp]
\includegraphics[width=3.2in,height=2.8in]{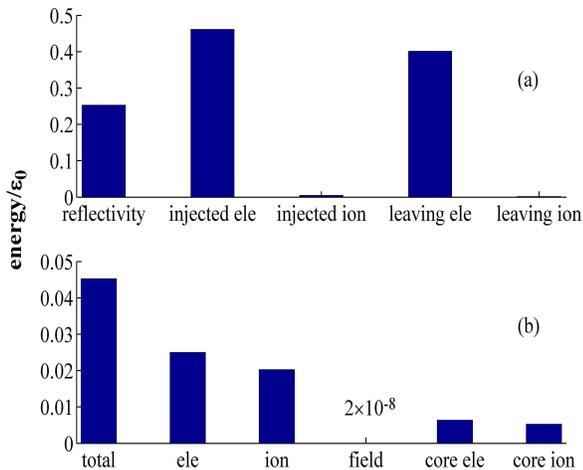}
\caption{\label{fig:epsart}(a) Energy of reflected light, the fast
electrons and ions injected to system II, and the fast electrons and
ions leaving the simulation box, respectively, normalized by the
incident laser energy $\varepsilon_0$. (b) Energy gain of system II,
background electrons and ions, fields, and the electrons and ions at
the core area, respectively, normalized by $\varepsilon_0$. These
values are obtained at 1.2 ps.}
\end{figure}

The bars in Fig. 10(a) show the ratios of the total energy of fast
particles and fields to the incident laser energy at 1.2 ps. The
reflected light carries off 25\% energy of the incident laser. The
conversion efficiency to fast particles is 46\% , though 87\% of
these leave the simulation box because of large divergence as well
as too high energy. Note that 0.47\% energy is absorbed by fast
ions. Figure 10(b) presents the energy gain partition of system II.
4.6\% energy of the incident laser is transferred to the background
plasma and the conversion efficiency to the core is only 1.2\%.
These values should grow slightly if a longer simulation is taken
since about 1.5\% laser energy is covered by the fast particles
which are still located in the system. It should be pointed out that
we have chosen too high laser intensity to examine our approach with
a powerful fast current. The optimized laser parameters for ignition
should be investigated further in the future.

\section{Summary and discussion}
In summary, we have proposed a two-system PIC approach with which
integrated simulation can be performed including fast particle
generation via laser plasma interaction and fast particle transport
and energy deposition in extremely high density plasma. This
approach can be applied to an integrated simulation of fast ignition
with real target density, e.g., 300 $\rm g/cm^3$, which has been
illustrated by the example in this paper. Also, it may be used to
simulate the transport of fast particles in solid targets with a
self-consistent calculation of the fast particle generation, e.g.,
in the context of hard x-ray sources. To apply this approach, a
Coulomb collision module as well as a high order current scheme
should be included in the PIC code.

In the two-system approach, a full PIC model with collisions is
taken in the first system, which is used to calculate the fast
particle generation. In the second system a reduced field solver is
employed, as used in hybrid PIC approach. This system computes fast
particle transport and energy deposition in the high density plasma.
The plasmas in both systems are modeled by macroparticles as in a
conventional PIC code. If the energies of particles generated in the
first system are above some threshold (e.g., hundreds of keV
adjustable according to different physical problems), their data
will be copied to the second system. The fast particles include not
only electrons but also ions and therefore, the two-system PIC model
could also be applied in ion transport and ion fast ignition.

In the current version, we have taken the same simulation box and
resolutions in the two systems for simplicity. In principle, one can
reduce the simulation box size in the full PIC system if there is a
large enough space between the injection point of the fast particles
and the right boundary of the simulation box, as shown in Appendix
C. Also, the two systems may have different resolutions in time and
space. For example, the resolutions in the second system can be
taken to be lower than the first system to save the computation
expense.

Basically, in the two-system model the hot particle transport is
simulated via a relay channel between the two PIC systems. The relay
point between the two systems is the injection point of the fast
particles. To assure the validation of the two-system model, the
relay point should be taken such that: it is far away from the laser
interaction zone; the plasma density at this point should be
sufficiently high enough to satisfy the fluid approximation applied
in the reduced field solver; the density should also be low enough
to reduce the noise in the full PIC system. In this paper we have
used a relay point with a density of 100 $n_c$. With a longer
duration, e.g., 10 ps, the laser pulse can enter into a deeper
region with a higher plasma density via the hole boring effect. The
relay point need be taken with an even higher density. This is the
case for FI with a cone-free target as used in our recent work
\cite{Wang_PRL}, in which the two-system approach has been applied
to study a new form of magnetically assisted FI.

\begin{acknowledgments}
W.M.W. acknowledges support from the Alexander von Humboldt
Foundation. This work was supported by the National Basic Research
Program of China (Grants No. 2013CBA01500) and NSFC (Grants No.
11375261, No. 11105217, No. 11129503, No. 113111048, No. 11375262,
and No. 11135012). The authors gratefully acknowledge the computing
time granted by the JARA-HPC and VSR committees on the supercomputer
JUQUEEN at Forschungszentrum J\"ulich.
\end{acknowledgments}

\appendix
\section{On the PIC code KLAPS}
The PIC code KLAPS (kinetic laser plasma simulation) has been
developed by Wei-Min Wang and Zheng-Ming Sheng from the 2D serial,
basic version of KLAP \cite{minchen} to a parallel code with both 2D
and 3D versions. Besides the basic properties of a traditional PIC
code \cite{PIC1,PIC2,Yee_solver}, KLAPS also includes field
ionization, Coulomb collision, radiation reaction, first-fourth
order ``zigzag'' current calculation, a Maxwell-equation-reduced
field solver, a dispersion-free field solver, moving window
technology, and absorbing and/or periodic boundary conditions in any
direction. In this paper, Coulomb collision, fourth order "zigzag''
current calculation, absorbing boundary conditions are applied in
the simulations.

\begin{figure}[htbp]
\includegraphics[width=3.2in,height=2.8in]{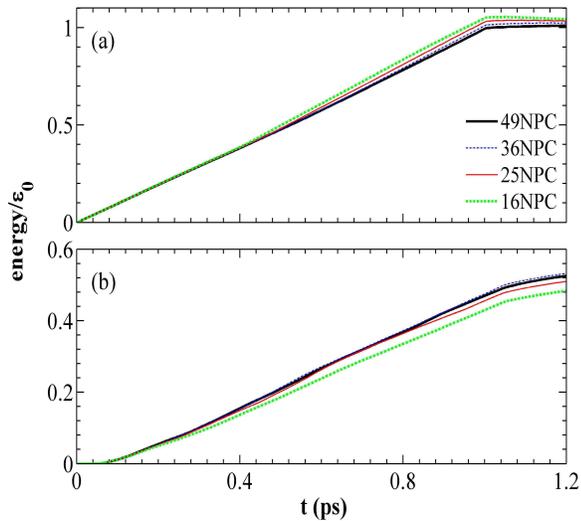}
\caption{\label{fig:epsart} (a) Evolution of the total energy gain
of system I with different particle numbers per cell (NPC) and (b)
the corresponding fast particle energy injected into system II,
where the energy is normalized by the total energy of the incident
laser pulse.}
\end{figure}

The Coulomb collision module has been developed based on the scheme
proposed by Sentoku and Kemp \cite{Sentoku_08}. This scheme computes
electron-electron, electron-ion and ion-ion collisions with a
relativistic formula of two-body collision
\cite{Sentoku_08,Sentoku_98}. Through a series of two-body collision
tests of randomly chosen pairs in a cell \cite{Collision1}, it can
simulate many-body collision and it can perfectly keep the
conservation of both energy and momentum per time step with
macroparticles of different weights \cite{Sentoku_08}. KLAPS
calculate currents by 1st-4th order ``zigzag'' scheme. The first
order algorithm proposed by Umeda \emph{et al.}
\cite{Current_zigzag} in 2003, which shows higher efficiency than
the one proposed by Esirkepov \cite{Current_Esirkepov}. We extended
it to 2nd-4th order with charge conservation
\cite{Current1,Current2} and implemented them in KLAPS. The field
ionization of atoms is calculated according to the
Ammosov-Delone-Krainov (ADK) formula \cite{ADK,ADK_revised}. Besides
tunneling ionization at high laser intensities with the Keldysh
parameter $\gamma_K<1$, the ADK model can also be applied in the
intermediate regime between multiphoton ionization and tunneling
with $1<\gamma_K<8$ at low laser intensities \cite{MultiPho_ADK},
where the Keldysh parameter is refined by
$\gamma_K=\sqrt{\Phi/2U_p}$ \cite{Keldysh}, $\Phi$ is the ionization
potential, and $U_p$ is the ponderomotive potential of the laser
pulses. With this module it is possible to simulate ionization at a
wide range of laser intensities, e.g., above $10^{13}~ \rm W/cm^2$.
We have applied this module to investigate terahertz (THz) radiation
generation from gas ionization
\cite{Wang1,Midinfared,Wang_pre,Wang_pra}. The radiation reaction
effect is added in the particle motion equation, where the radiation
damping force is calculated according to the Landau-Lifshitz formula
\cite{Landau-Lifshitz}. With this module the code has the potential
to simulate laser plasma interactions with the intensity up to
$10^{22}~ \rm W/cm^2$, above which weak quantum electrodynamics
effects starts to work.

\section{Requirement of resolutions}

\begin{figure}[htbp]
\includegraphics[width=3.2in,height=2.8in]{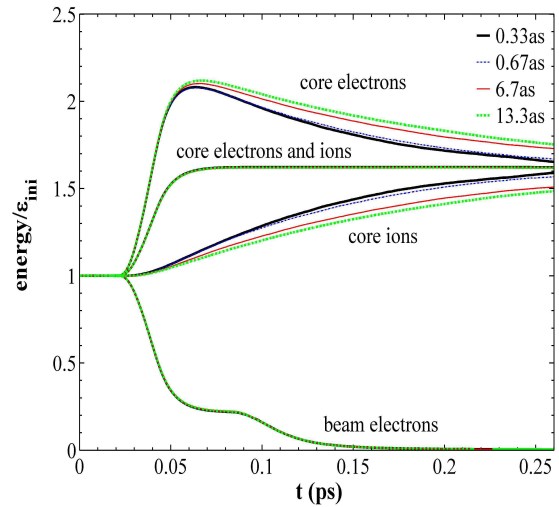}
\caption{\label{fig:epsart}Evolution of the average energy of the
core electrons and ions as well as fast electrons with different
collision time steps, where the energy units are taken as the
initial values $\varepsilon_{ini}$ of each species.}
\end{figure}

We present simulations to examine the requirement of resolutions in
the two systems, respectively. The density setup in the two systems
is taken the same as shown in Fig. 2. The cone angle and tip size is
changed to 20.6 degree and 2 $\rm\mu m$. The same laser parameters
are employed except a smaller spot radius of $r_0=2~\rm\mu m$. The
simulation size $40\rm\mu m \times 32\rm\mu m$ is taken in $x\times
y$ directions.

The requirements of the first system are mainly limited by the
numerical noise caused by unresolved plasma oscillation which is
characterized by the plasma frequency $\omega_p$ and skin depth
$c/\omega_p$. Hence, one usually need take the spatial cell size
$\Delta x$ and the temporal step $\Delta t$ to be comparable with or
smaller than $c/\omega_p$ and $1/\omega_p$, respectively. These
restraints can be relaxed with a high order interpolation. We take
the cone wall density of 100 $n_c$ and the highest density is also
this value in system I. By use of the 4th order ``zigzag'' scheme it
is shown that the noise can be controlled well with a spatial size
$\Delta x=\Delta y = (1/32)\mu m=1.96c/\omega_p$ and the time step
$\Delta t=0.64\Delta x/c=1.25/\omega_p$ if the particle number per
cell (NPC) is larger than 16, as shown in Fig. 11(a). This figure
illustrates the evolution of the system energy normalized by the
total energy of the incident laser pulse. At 1.2 ps the errors
caused by the noise are 0.94\%, 2.3\%, 3.2\%, 4.3\%, respectively,
for 49 NPC, 36 NPC, 25 NPC, and 16 NPC.

Figure 11 (b) plots the fast particle energy injected to system II.
We change NPC to check the anomalous macroparticle stopping effect,
which limits the highest number of real particles denoting by a
macroparticle or the weight of a macroparticle. The evolution curve
with 36 NPC is nearly coincident with the one with 49 NPC and its
error ratio to the latter is 1.3\% at 1.2 ps. The errors are 4.3\%
and 9.2\% for the cases with 25 NPC and 16 NPC. Summing up the two
examination results given in Figs. 11(a) and 11(b), 25 NPC is good
enough to controlled both the noise and the anomalous stopping
effect for system I.

\begin{figure}[htbp]
\includegraphics[width=3.2in,height=2.8in]{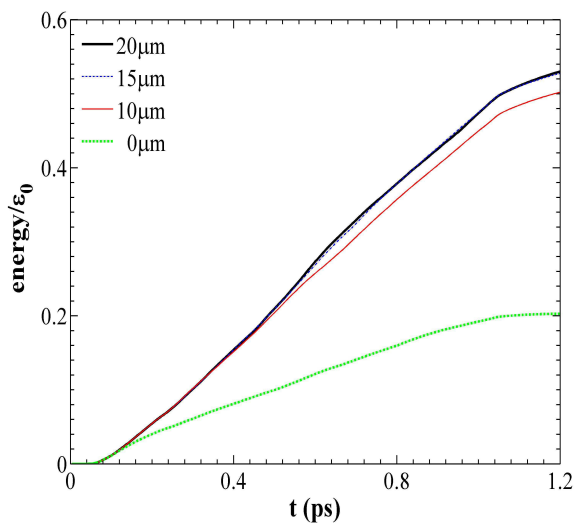}
\caption{\label{fig:epsart} Fast particle energy injected into
system II, where it is normalized by the total energy of the
incident laser $\varepsilon_0$. Different curves correspond to the
data of the fast particles deleted at 0 $\mu m$, 10 $\mu m$, 15 $\mu
m$, 20 $\mu m$ away from the injection point in system I. }
\end{figure}

In system II the reduced field solver is applied which has much
lower resolution requirement than the full solver since the
fast-varied displacement current is omitted. Simply, one can safely
take the same resolutions as system I. The bottleneck of the
temporal resolution is imposed by collisions. One need make sure
that the time step of the collision $\Delta t_c$ is smaller than the
collision period $\tau_c=1/\nu_c$ (the principle to choose $\Delta
t_c$ can be found in Refs. \cite{Sentoku_08,Sentoku_98}), where
$\nu_c=4\pi e^2 n_e L/(p_{rel}^2v_{rel})$ is the relativistic,
two-body collision frequency of a particle pair \cite{Sentoku_08},
$L$ is the Coulomb logarithm, as well as $p_{rel}$ and $v_{rel}$ the
relative momentum and velocity of a particle in the rest reference
of the other. We evaluate $\tau_c$ with the core density of 300
$g/cm^3=54000~n_c$ and temperature of 1 keV. For simplicity we
compute $p_{rel}$ with the assumption of the core particle at rest.
For the collision between the core electron and fast electrons of 1
MeV, $\tau_c=$ 424 fs. This value is 19 fs, 0.93 fs, respectively,
for fast electrons of 0.1 MeV, 0.01 MeV. The period of collision
between a core electron and ion is much smaller, which is 59 as with
the initial temperature 1 keV. We take a small enough $\Delta t_c=$
0.33 as to resolve all the collision periods as the standard result.
We also take $\Delta t_c$ between 0.67 as and 13.3 as to compare
with the standard result. To only check the collision effect, we
take all of the electrons in the cone with monoenergetic energy of
0.3 MeV and the motion along +x direction and no laser pulse
incident. One can see in Fig. 12 that the evolution of the fast
electrons as well as the whole core electrons and ions nearly
coincide when $\Delta t_c$ is changed, since all the $\Delta t_c$
used are much smaller than the collision period between the fast
electrons and the core. The gain of the core ion temperature is 0.6
keV with $\Delta t_c=$ 0.33 as at the time of 0.27 ps. From the
value, the deviation of the gain with $\Delta t_c=$ 0.67 as, 6.7 as,
and 13.3 as, respectively, is 3\%, 13\% and 16\%. Therefore, $\Delta
t_c=$ 0.67 as is good enough. Considering that the collision period
will increase with the enhancing core temperatures, one could use a
larger $\Delta t_c$. If one focuses on the energy transfer from the
fast electrons to the core, $\Delta t_c$ can be taken to be much
larger, e.g., $\Delta t_c=$ 13.3 as.

\section{Comparison between particle data duplication and migration}

In all of the previous simulations, we have copied fast particle
data from system I to II and meanwhile these fast particles still
exist in system I. If the data are deleted from system I just after
transferred to system II, the computation expense can be saved.
However, an unphysical electric field will be formed around the
injection point in system I to block the coming particles. This
effect on the particle injection is shown in Fig. 13. If the fast
particle data deleted in system I immediately at the injection point
after they transferred, the injection energy is reduced
considerably. However, if the fast particle data deleted far away
from the injection point, i.e., above 15 $\mu m$, the injection
electron energy is affected slightly.

\section{Benchmark of two-system PIC}


Figure 14 shows the benchmark of the two-system PIC model against
the full and two-boundary PIC models, where the results of the full
and two-boundary PIC simulations have been presented in Fig. 1.
Basically the field solver used in the second system of our
two-system PIC model is nearly the same as the one taken in the high
density region in the two-region PIC, as addressed in Sec. III.
Therefore, the two simulation results show good agreement as seen in
Fig. 14. Here, we adopt a decreased spatial resolution of 20 nm in
the two-system simulation while it is 5 nm in the other two
simulations.


\end{document}